\documentclass[aps,twocolumn,superscriptaddress,showpacs,nofootinbib]{revtex4-1}

\usepackage[dvipsnames]{xcolor}
\usepackage{graphicx}
\usepackage[subfigure]{graphfig}
\usepackage{epsfig}
\usepackage{dcolumn}
\usepackage{amsmath}
\usepackage{multirow}
\usepackage{braket}

\def\bal#1\eal{\begin{align}#1\end{align}}

\newcommand{\cpt}{Aix-Marseille Université, Université de Toulon, CNRS, CPT, Marseille, France}
\newcommand{\uky}{Department of Physics and Astronomy, University of Kentucky, Lexington, KY 40506, USA}
\newcommand{\ucas}{University of Chinese Academy of Sciences, School of Physical Sciences, Beijing 100049, China}
\newcommand{\cas}{CAS Key Laboratory of Theoretical Physics, Institute of Theoretical Physics, Chinese Academy of Sciences, Beijing 100190, China}
\newcommand{\sfp}{School of Fundamental Physics and Mathematical Sciences, Hangzhou Institute for Advanced Study, UCAS, Hangzhou 310024, China}
\newcommand{\ict}{International Centre for Theoretical Physics Asia-Pacific, Beijing/Hangzhou, China}

\def\path{./figures/}

\begin{document}

\title{\vspace{1.0in}Muon g\,-\,2 with overlap valence fermions
\vspace*{-0.5cm}
\begin{center}
\large{
\vspace*{0.4cm}
\includegraphics[scale=0.20]{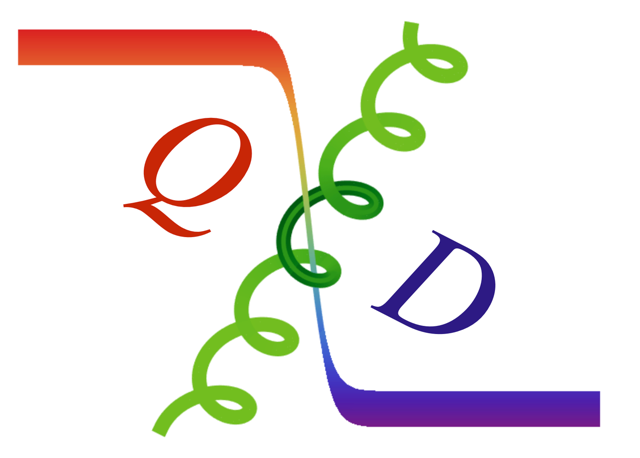}\\
\vspace*{0.4cm}
($\chi$QCD Collaboration)
}
\end{center}
}

\author{Gen Wang}\email{gen.wang@univ-amu.fr}\affiliation{\cpt}\affiliation{\uky}
\author{Terrence Draper}\affiliation{\uky}
\author{Keh-Fei Liu}\affiliation{\uky}
\author{Yi-Bo Yang}\email{ybyang@itp.ac.cn} \affiliation{\cas}  \affiliation{\ucas} \affiliation{\sfp}\affiliation{\ict}

\begin{abstract}
{
We present a lattice calculation of the leading order (LO) hadronic vacuum polarization (HVP) contribution to the muon anomalous magnetic moment for the connected light and strange quarks, $a^{\rm W}_{{\rm con}, l/s}$ in the widely used window $t_0=0.4~\mathrm{fm}$, $t_1=1.0~\mathrm{fm}$, $\Delta=0.15~\mathrm{fm}$, and also of $a^{\rm S}_{{\rm con}, l/s}$ in the short distance region.
We use overlap fermions on 4 physical-point ensembles.
Two 2+1 flavor RBC/UKQCD ensembles use domain wall fermions (DWF) and Iwasaki gauge actions at $a = 0.084$ and 0.114 fm, and two 2+1+1 flavor MILC ensembles use the highly improved staggered quark (HISQ) and Symanzik gauge actions at $a = 0.088$ and 0.121 fm.
We have incorporated infinite volume corrections from 3 additional DWF ensembles at ${\rm L}$ = 4.8, 6.4 and 9.6 fm and physical pion mass. 
For $a^{\rm W}_{{\rm con}, l}$, we find that our results on the two smaller lattice spacings are consistent with those using the unitary setup, but those at the two coarser lattice spacings are slightly different. Eventually, we predict $a^{\rm W}_{{\rm con}, l}=206.7(1.5)(1.0)$ and $a^{\rm W}_{{\rm con}, s}=26.8(0.1)(0.3)$, using linear extrapolation in $a^2$, with systematic uncertainties estimated from the difference of the central values from the RBC/UKQCD and MILC ensembles.
}

\end{abstract}

\maketitle

\section{Introduction}

The anomalous magnetic moment of the muon ($a_{\mu}\equiv (g_{\mu}-2)/2$) is one of the crucial benchmarks to verify the correctness of the standard model. The current analysis of the Fermilab experiment~\cite{Muong-2:2021ojo, Muong-2:2021vma} is consistent with the previous BNL E821~\cite{Muong-2:2006rrc} result with comparable precision, and
the Fermilab experiment is planned to reduce the uncertainty by a factor of 4. Those results are higher than the current standard model predictions~\cite{Aoyama:2020ynm} using phenomenological estimates by around 4$\sigma$, and so have attracted much theoretical interest about possible new physics. 

But such a deviation is very sensitive to the theoretical prediction of the strong interaction contribution to $a_{\mu}$, especially the leading order hadronic vacuum polarization (LO-HVP) contribution, $a_{\mu}^{\rm LO-HVP}$. The most recent determinations from the
hadronic $R$-ratio, a dispersion integral over hadronic cross section ratio $\sigma(e^+ e^- \rightarrow {\rm{hadrons}}) / \sigma(e^+ e^- \rightarrow \mu^+ \mu^-)$, are $\bar{a}\equiv a_{\mu}^{\rm LO-HVP}\times 10^{10}= 693.9(4.0)$~\cite{Davier:2019can} and 692.8(2.4)~\cite{Keshavarzi:2019abf},
while that required for no  new physics is 718(4)~\cite{Muong-2:2021ojo,Muong-2:2021vma,Muong-2:2006rrc, Aoyama:2020ynm}. Compared to $a_{\mu}^{\rm LO-HVP}$, the electron mass suppresses $a_{e}^{\rm LO-HVP}$, the corresponding quantity for the electron, by a factor of $(m_e/m_{\mu})^2\sim 10^{-4}$, and so the theoretical uncertainty of HVP will not affect the agreement between the current theoretical and experimental determinations of $a_e$.

$a_{\mu}^{\rm LO-HVP}$ can be obtained using first-principles Lattice QCD calculations, which avoid the possible phenomenological uncertainty from the $R$-ratio determination. There are many recent Lattice QCD results~\cite{Budapest-Marseille-Wuppertal:2017okr,RBC:2018dos,Gerardin:2019rua,FermilabLattice:2019ugu,Giusti:2019xct,Borsanyi:2020mff,Aubin:2022hgm,Ce:2022kxy,FermilabLattice:2022smb,Alexandrou:2022amy,RBCUKQCD:2022,Blum:2023qou}, and the most precise one~\cite{Borsanyi:2020mff} obtains $\bar{a}=707.5(5.5)$ and agrees with the ``no new physics" requirement within 1.5$\sigma$. 
Recent studies~\cite{Aubin:2019usy,RBCUKQCD:2022,Blum:2023qou} suggest that the so-called ``window'' value~\cite{RBC:2018dos} of $a_{\mu}$, which picks the contribution around the $\rho$ meson pole, can be sensitive to the discretized fermion action if the continuum extrapolation is not conservative enough.
In this work, we will calculate the window value of $a_{\mu}$ using the overlap valence fermion action on ensembles with either the Domain Wall fermion (DWF) sea from the RBC/UKQCD collaboration~\cite{RBC:2012cbl,Blum:2014tka,Boyle:2015exm} or the HISQ sea from the MILC collaboration~\cite{Bazavov:2012xda, Bazavov:2017lyh}, to study the fermion action dependence. We will also examine the lattice spacing dependence in these cases. 

The numerical setups of this work are collected in Sec.~\ref{sec:setup}. Section~\ref{sec:result} presents our results with their statistical and systematic uncertainties. The summary and extended discussions are given in Sec.~\ref{sec:summary}.

\section{Numerical setup}\label{sec:setup}

One can calculate $a_{\mu}^{\rm LO-HVP}$ with the following expression,
\bal
a_{\mu}^{\rm LO-HVP}&=4\alpha^2 \int_0^{\infty} \frac{\mathrm{d}q^2}{m^2_{\mu}}  f(\frac{q^2}{m^2_{\mu}}) (\Pi(q^2)-\Pi(0)),\\
f(r)&=\frac{Z^3(r)(1-\sqrt{r}Z(r))}{\sqrt{r}(1+Z^2(r))},\ Z(r)=\frac{\sqrt{r+4}-\sqrt{r}}{2},\nonumber
\eal
where $\alpha=\frac{e^2}{4\pi}\sim 1/137$ is the fine structure constant, $m_{\mu}$ is the muon mass, the HVP $\Pi(q^2)$ can be obtained from the Fourier transform of the vector current two-point function in Euclidean space-time, 
\bal
\Pi^{\mu\nu}(q)=\int \mathrm{d}^4 x e^{{\rm i}q x} \langle j_{\mu}(x)j_{\nu}(0) \rangle=\Pi(q^2)(q^2\delta_{\mu\nu}-q_{\mu}q_{\nu}),
\eal
and the electromagnetic current $j_{\mu}=\sum_f Q_f \bar{\psi}_f\gamma_{\mu}\psi_f$ is summed over all the quark flavors $f=u,d,s,c,...$ with their electric charge $Q_f$ in units of the electron charge $e$. The factor $Z$ has the properties $Z(0)=1$ and $Z(r)\propto 1/\sqrt{r}$ at $r\rightarrow \infty$, which ensures that the major contribution in the $q^2$ integration comes from the small $q^2$ region.

One can select the momentum $q$ to be along the temporal direction to simplify the expression of $\Pi(q^2)$ to
\bal
\Pi(q^2)=\int \mathrm{d} t \frac{\mathrm{cos}(t q)}{q^2}C(t),
\eal
where $C(t)\equiv \frac{1}{3}\sum_i\langle \int \mathrm{d}^3 x \, j_i(\vec{x},t)j_i(\vec{0},0)\rangle$. This definition includes the $1/q^2$ divergence, but then the subtracted $\Pi(q^2)$~\cite{Bernecker:2011gh} is
\bal
\Pi(q^2)-\Pi(0)=\int \mathrm{d} t \left[\frac{\mathrm{cos}(t q)-1}{q^2}+\frac{1}{2}t^2\right]C(t).
\eal
One can show that $\Pi(q^2)\propto 1/M^2$ if $C(t)\propto e^{-Mt}$ and then $a_{\mu}^{\rm HVP}\propto m_{\mu}^2/M^2$ if $C(t)$ is dominated by a single state with mass $M\gg m_{\mu}$, and then the heavy quark contribution to $a_{\mu}^{\rm HVP}$ is also suppressed by $1/m_Q^2$.

Eventually one can rewrite $a_{\mu}^{\rm LO-HVP}$ in terms of $C(t)$ and the weight function $\omega(t)$,
\bal
&a_{\mu}^{\rm LO-HVP}=\int \mathrm{d} t\,\omega(t) C(t),\nonumber\\
&\quad \omega(t)=4\alpha^2\int_0^{\infty} \frac{\mathrm{d}q^2}{m^2_{\mu}}  f\left(\frac{q^2}{m^2_{\mu}}\right) \left[\frac{\mathrm{cos}(t q)-1}{q^2}+\frac{1}{2}t^2\right]\label{eq:def_omega}.
\eal
On the lattice, one could use the original $\omega(t)$ or replace $\omega(t)$ by its lattice version~\cite{RBC:2018dos},
\bal\label{eq:def_omega2}
\hat{\omega}(t)=4\alpha^2\int_0^{\infty}\frac{\mathrm{d}q^2}{m^2_{\mu}}  f\left(\frac{q^2}{m^2_{\mu}}\right) \left[\frac{\mathrm{cos}(t q)-1}{[\frac{2}{a}\mathrm{sin}(\frac{qa}{2})]^2}+\frac{1}{2}t^2\right],
\eal
and sum $t$ over all the discretized time slices $(-\frac{Ta}{2},\frac{Ta}{2}]$:
\bal
a_{\mu}^{\rm LO-HVP, \omega}&= \left. \left[ \sum \omega(t) C^{\rm lat}(t,a) \right] \right|_{a\rightarrow 0,T\rightarrow \infty},\\
a_{\mu}^{\rm LO-HVP, \hat{\omega}}&=\left. \left[\sum \hat{\omega}(t) C^{\rm lat}(t,a)\right] \right|_{a\rightarrow 0,T\rightarrow \infty},
\eal 
 to obtain the final prediction of $a_{\mu}^{\rm LO-HVP}$, 
where $a$ is the lattice spacing, $T$ is the dimensionless number of lattice sites along the temporal direction, and
$C^{\rm lat}(t,a)=a\big\{C(t)+{\cal O}(a^2)\big\}$ is the correlation function on the lattice. 
Based on numerical calculation, changing the upper limit of the integral in Eq.~\ref{eq:def_omega2} to a finite constant, such as $(\pi/a)^2$, changes the value of the integral by less than 0.005\%, which is much smaller than our other uncertainties.
Also, Ref.~\cite{Budapest-Marseille-Wuppertal:2017okr} suggests that the correction is about $0.02\%$, by introducing an integral cut-off $Q_{\rm{max}}^2 = 3 \ {\rm{GeV}}^2$ in Eq.~\ref{eq:def_omega2}, which is also much smaller than all our total uncertainties.

In the practical calculation on the ensemble with 2 degenerate light flavors, $C(t)$ can be decomposed into several pieces~\cite{RBC:2018dos}, 
\bal
C(t)&=\frac{5}{9}C^{\rm con}(t;m_l)+\frac{1}{9}C^{\rm con}(t;m_s)+\frac{4}{9}C^{\rm con}(t;m_c)\nonumber\\
&\quad +C^{\rm dis}(t)+\alpha C^{\rm QED}(t)+\Delta m C^{\rm SIB}(t)\nonumber\\
&\quad+{\cal O}(\alpha^2,\alpha\Delta m, \Delta m^2),
\eal
where $m_{l}=(m_d+m_u)/2$ is the iso-symmetric light quark mass, $m_{s,c}$ are the strange and charm quark masses respectively, $C^{\rm con}(t;m_f)=C^{\rm con}(t,S(\vec{x}=\vec{0},t_0=0\,;m_f))$,
\bal
C^{\rm con}(t,S(\vec{x}_0,t_0;m_f))&=\frac{1}{3}\sum_i\langle \{\int \mathrm{d}^3 x \mathrm{Tr}[\gamma_i\gamma_5\nonumber\\
&\quad \quad S^{\dagger}(\vec{x},t+t_0;\vec{x}_0,t_0;m_f)\gamma_5\gamma_i\nonumber\\
&\quad \quad S(\vec{x},t+t_0;\vec{x}_0,t_0;m_f)]\rangle
\eal
is the connected correlation function with the quark propagator $S(m)\equiv 1/(D+m)$, $C^{\rm dis}$ is the contribution from the disconnected quark diagram, $C^{\rm QED}$ is the leading order QED correction which can be accessed from the 4-point correlation function with an infinite-volume photon~\cite{Lehner:2015bga} (the next order contribution is negligible for the precision required by $a_{\mu}$), and $C^{\rm SIB}$ is the strong isospin breaking (SIB) effect which is proportional to $(m_d-m_u)/2$. 

In this work, we use overlap fermions on several gauge ensembles with 1-step HYP smearing to calculate $a^{\rm W}_{\mu}$. The ensembles include the 2+1 flavor DWF ensembles with the Iwasaki gauge action from the RBC/UKQCD collaboration~\cite{RBC:2012cbl,Blum:2014tka,Boyle:2015exm} and the 2+1+1 flavor HISQ ensemble with the Symanzik gauge action from the MILC collaboration~\cite{Bazavov:2012xda, Bazavov:2017lyh}.

The overlap fermion action uses a matrix sign function $\epsilon(H_{\rm w})=\frac{H_{\rm w}}{\sqrt{H_{\rm w}^2}}$ of the hermitian Wilson Dirac operator $H_{\rm w}(-M_0)=\gamma_5D_{\rm w}(-M_0)$ to construct
\begin{align}
D_{\textrm{ov}}=M_0\Big(1+\gamma_5\epsilon\big(H_{\rm w}(-M_0)\big)\Big),
\end{align}
 which was proposed in Ref.~\cite{Chiu:1998eu,Liu:2002qu} as a discretized fermion operator satisfying the Ginsburg-Wilson relation $D_{\textrm{ov}}\gamma_5+\gamma_5D_{\textrm{ov}}=\frac{a}{M_0}D_{\textrm{ov}}\gamma_5D_{\textrm{ov}}$~\cite{Ginsparg:1981bj}, where $D_{\rm w}$ is the Wilson Dirac operator with a negative mass such as $M_0=1.5$. The DWF action can be considered as an approximation of the overlap fermion action using a slightly inaccurate $\epsilon(x)$, and the HISQ action~\cite{Follana:2006rc} is a modified version of the staggered fermion action which is much cheaper than either DWF or overlap but which suffers from the taste mixing effect (see Ref.~\cite{Aubin:2019usy}, for example).  
 
Using the overlap fermion action for the valence quark allows us to implement low-mode substitution~\cite{xQCD:2010pnl,XQCD:2013odc} to improve the signal based on the low-lying eigenvectors of $D_c$, 
\bal
C^{\rm con, LMS}(t)=&\frac{1}{N_{\rm src}N_g^3}\sum_i\big\{C^{\rm con}(t,S^{\rm grid}(\vec{x}_i,t_i))\nonumber\\
&- C^{\rm con}(t,S^{\rm grid}_{L}(\vec{x}_i,t_i))\big\}\nonumber\\
&+\frac{1}{L^3T}\sum_{\vec{x},t}C^{\rm con}(t,S_L(\vec{x},t)),
\eal 
where $N_{\rm src}$ is the number of $S^{\rm grid}$ located at different origins $(\vec{x}_i,t_i)$, $S_L(m)\equiv \sum_{|\lambda|<\lambda_c} \frac{1}{\lambda+m}v_{\lambda}v_{\lambda}^{\dagger}$,  $v_{\lambda}$ satisfies $D_cv_{\lambda}=\lambda v_{\lambda}$ and $\lambda_c \sim $ 200 MeV is the upper bound of the eigenvalue $\lambda$. The quark propagator $S^{\rm grid}(\vec{x}_i,t_i)=\sum_{y\in{\rm grid}}S(\vec{x},t+t_0;\vec{x}_i+\vec{y},t_i)$ above uses a random ${\rm{Z}}_3$~\cite{Dong:1993pk} grid source with $N_g$ points in each spacial dimension. Such a grid source has a starting point $\vec{x}_0 = (x_0,y_0,z_0)$ and 
$\vec{x}_i \in (x_0 + m_x \Delta_x, y_0 + m_y \Delta_y,z_0 + m_z \Delta_z)$
where $\Delta_{x,y,z} = L/N_g$ is the spatial direction offset
 and $m_{x,y,z} \in \{0,1,\cdots, L/ \Delta_{x,y,z}\}$ is the offset number in each direction for each grid point, and $n= N_g^3$ is the number of grid points of the grid source.
 
 In order to evaluate the standard window with $t_0=0.4~\mathrm{fm}$, $t_1=1.0~\mathrm{fm}$, $\Delta=0.15~\mathrm{fm}$ efficiently, we have chosen $\Delta_{x,y,z} \sim 1.0 \ {\rm{fm}}$ to reduce the number of inversions needed. The low-mode source point $\vec{x},t$ loops over the whole lattice volume to have full statistics for the low-mode parts. The information of the gauge ensembles, grid source parameters, and $\lambda_c$ are listed in Table~\ref{tab:ensemble}.
 
\begin{table}[ht!]                   
\caption{Information of the ensembles, grid sources and $\lambda_c$ used in this calculation, including the DWF+Iwasaki ensembles 48I and 64I~\cite{Blum:2014tka,Boyle:2015exm} and also the HISQ+Symanzik ensembles a12m130 and a09m130~\cite{Bazavov:2017lyh} with the lattice spacings from Ref.~\cite{Aubin:2019usy}. Three DWF+Iwasaki+DSDR ensembles 24D/32D/48D~\cite{RBC:2012cbl,Boyle:2015exm} are used to estimate the finite-volume effect, and four HISQ+Symanzik ensembles a04m310/a06m310/a09m310/a12m310~\cite{Bazavov:2012xda, Bazavov:2017lyh} with the lattice spacings from Ref.~\cite{Zhang:2020rsx} are used to study the continuum extrapolation. The pion mass $m_{\pi}$ and upper bound $\lambda_c$ of the eigenvalues are in the unit of MeV. }  
\begin{tabular}{ l l l c l c c c}                                                                
 \text{Symbol} & $L^3 \times T$  &  $a$ (fm)   & $m_{\pi}$   &  $N_\text{cfg}$ & $N_{\rm src}$ & $N_g$ & $\lambda_c$\\
\hline   
 48I      &$48^3\times\ 96$&0.11406(26) & 139  & 100 & 12 & 4 & 234 \\  
 64I      &$64^3\times128$ & 0.08365(25) & 139  &  92 &  8 & 4 & 187 \\    
\hline
 a12m130  &$48^3\times\ 64$& 0.12121(64) & 131 &  23 & 8 & 4 & 180 \\  
 a09m130  &$64^3\times\ 96$& 0.08786(47) & 128 &  22 & 8 & 4 & 200 \\   
 a12m310 &$24^3\times\ 64$&0.12129(89)   & 305 & 54  & 16& 1 & 224 \\  
 a09m310  &$32^3\times\ 96$& 0.08821(71) & 313 & 39  & 16& 1 & 195 \\
 a06m310  &$48^3\times 144$& 0.05740(50) & 319 & 32  & 8 & 1 & 243 \\  
 a04m310  &$64^3\times 192$& 0.04250(40) & 310 & 54  & 2 & 1 & 167 \\ 
\hline
 24D  &$24^3\times\ 64$& 0.1940(19) & 141 &  232 & 8 & 2 & 263 \\  
 32D  &$32^3\times\ 64$& 0.1940(19) & 141 &  134 & 8 & 4 & 230 \\  
 48D  &$48^3\times\ 64$& 0.1940(19) & 141 &  47  & 8 & 6 & 116 \\  
\hline
\end{tabular}  
\label{tab:ensemble}                                                                                  
\end{table}

\section{Results and systematics}\label{sec:result}

The window method proposed in Ref.~\cite{Bernecker:2011gh,RBC:2018dos} allows a more precise prediction to combine the ``window'' value of $a^{\rm LO-HYP}_{\mu}$ using the $C(t)$ from lattice QCD
\bal
&a^{\rm W}_{\mu}=\int \mathrm{d} t\,w^{\rm W}(t)\omega(t) C(t),\\
&w^{\rm W}(t)\equiv\theta(t,t_0,\Delta)-\theta(t,t_1,\Delta),\\
&\quad \theta(t,t',\Delta)\equiv\frac{1}{2}\big(1+\mathrm{tanh}[(t-t')/\Delta]\big),
\eal
with the remaining parts 
\begin{flalign}
&a^{\rm etc.}_{\mu}=a^{\rm S}_{\mu}+a^{\rm L}_{\mu},\nonumber\\
&a^{\rm S}_{\mu}=\int \mathrm{d} t\,w^{\rm S}(t)\omega(t) C(t), \\
&a^{\rm L}_{\mu}=\int \mathrm{d} t\,w^{\rm L}(t)\omega(t) C(t), 
\end{flalign}
with $w^{\rm S}(t)\equiv 1-\theta(t,t_0,\Delta)$, $w^{\rm L}(t)\equiv \theta(t,t_1,\Delta)$ and the $C(t)$ from the $R$-ratio.
The extra weight functions $w^{\rm S,W,L}$ pick the short, medium, and long distance contributions of $C(t)$, respectively, and separate $a_{\mu}$ into three pieces. With typical parameters ($t_0=0.4~\mathrm{fm}$, $t_1=1.0~\mathrm{fm}$, $\Delta=0.15~\mathrm{fm}$), $a^{\rm W}_{\mu}$ suppresses the contribution of $C(t)$ from $t\ll t_0$ and $t\gg t_1$, and can have smaller uncertainty using $C(t)$ from lattice QCD compared to that from the $R$-ratio.
$a^{\rm W}_{\mu}$ also provides a good reference to compare the independent lattice QCD results with good precision, in order to check the systematic uncertainties due to different lattice actions and their respective discretization errors. 

\begin{figure}[t]
    \centering
   \includegraphics[width=1\linewidth]{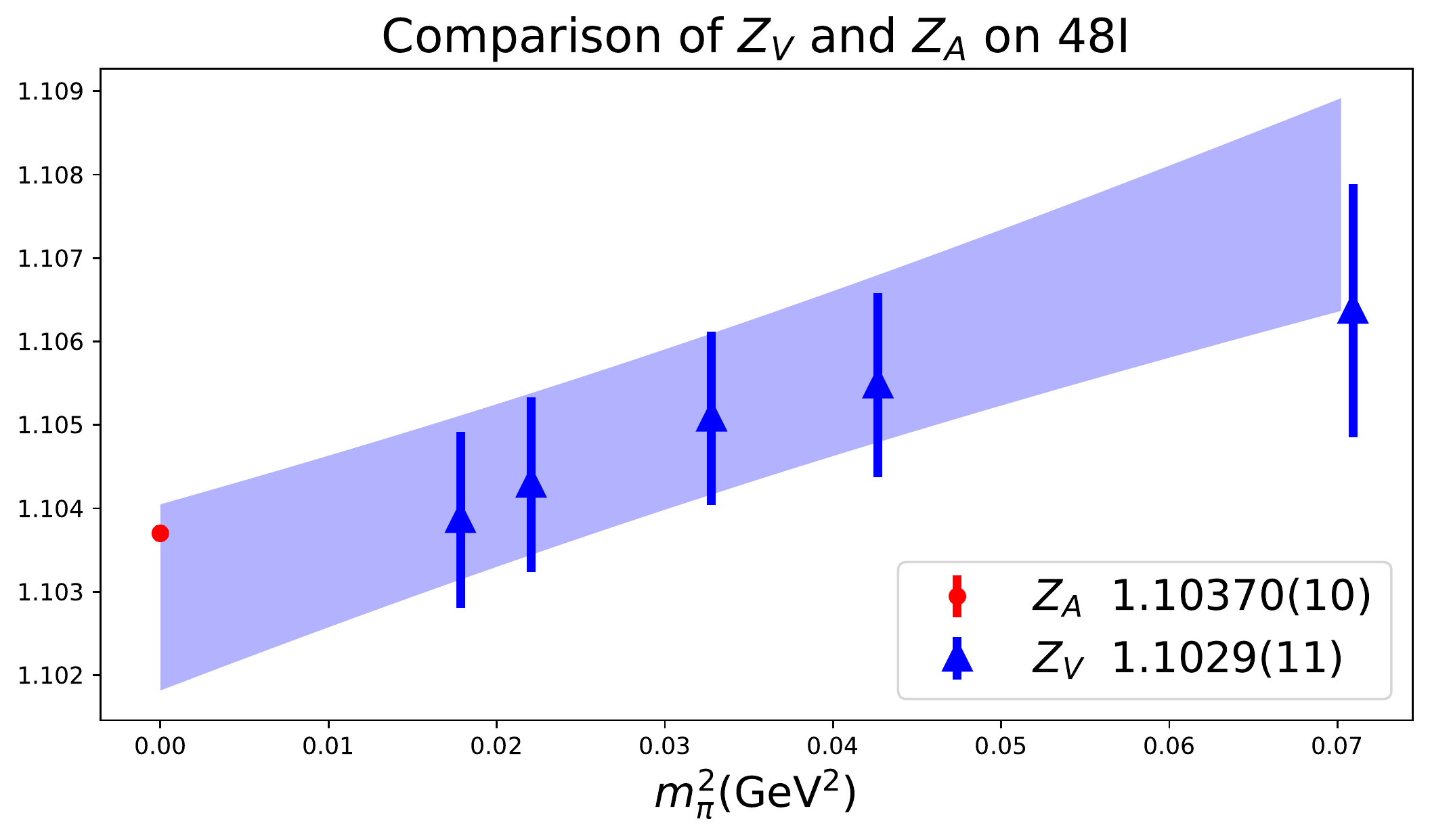}
  \caption{Comparison of the axial normalization constant and the vector normalization constant for the local vector current on 48I. The blue band is a linear fit of the vector normalization constant versus the pion mass squared, $m_\pi^2$. }
    \label{fig:ZA_ZV_48I}
\end{figure}

We shall define the rescaled connected light and strange quark contributions as
\bal
&\bar{a}^{\rm X}_{{\rm con},l}=\bar{a}^{\rm X}_{\rm con}(m_l),\ \bar{a}^{\rm X}_{{\rm con}, s}=\frac{1}{5}\bar{a}^{\rm X}_{\rm con}(m_s),\nonumber\\
&\bar{a}^{\rm X}_{\rm con}(m_q)\equiv \frac{5}{9}\int \mathrm{d} t\, w^{\rm X}(t)\omega(t) C^{\rm con}_f(t;m_q)\times 10^{10},
\eal
where ${\rm X\in \{S,W,L}\}$
and $m_l$ and $m_s$ are the physical light and strange quark masses, respectively. We use the local vector current in the calculation and apply the axial-vector normalization constant $Z_A$ obtained from PCAC~\cite{He:2021tve} since the local vector current normalization constant obeys $Z_V = Z_A$ for overlap fermions.
As shown in Fig.~\ref{fig:ZA_ZV_48I}, $Z_A$ agrees with $Z_V$ very well at the massless limit. ($Z_V$ is determined from the forward matrix element $Z_V \equiv {2 E}/{\bra{\pi(p)} V_4 \ket{\pi(p)}}$ with $\vec{p}=0$ in Ref.~\cite{Wang:2020nbf}.)
The uncertainty of $Z_A$ is at the 0.01\% level and can be ignored based on the precision target. Note that the last systematic uncertainty of $Z_A$ in Ref.~\cite{He:2021tve} is not necessary here as we don't need to extrapolate the strange quark mass in the sea to the chiral limit.

\begin{figure}[t]
    \centering
    \includegraphics[width=1\linewidth]{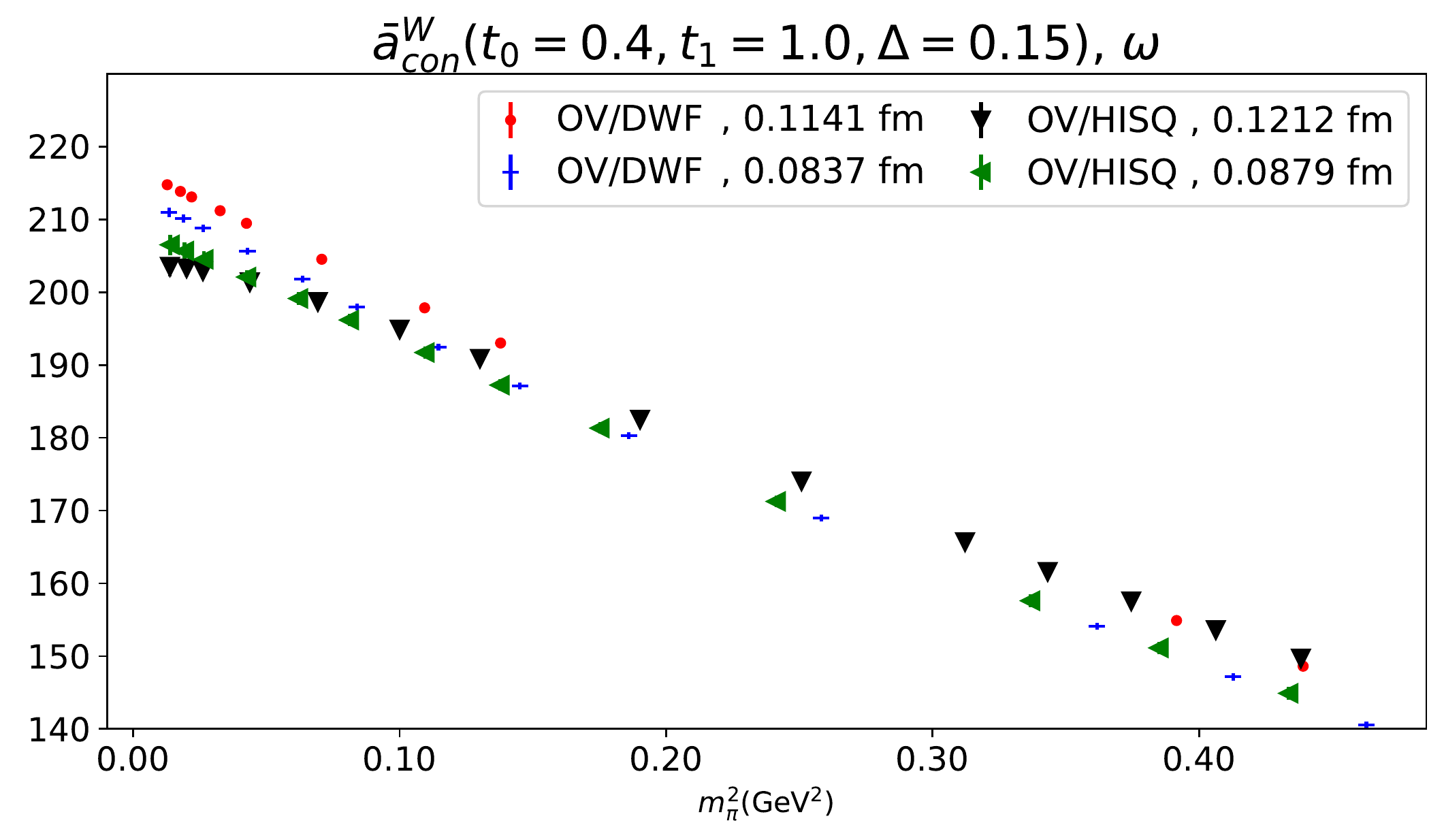}
   \includegraphics[width=1\linewidth]{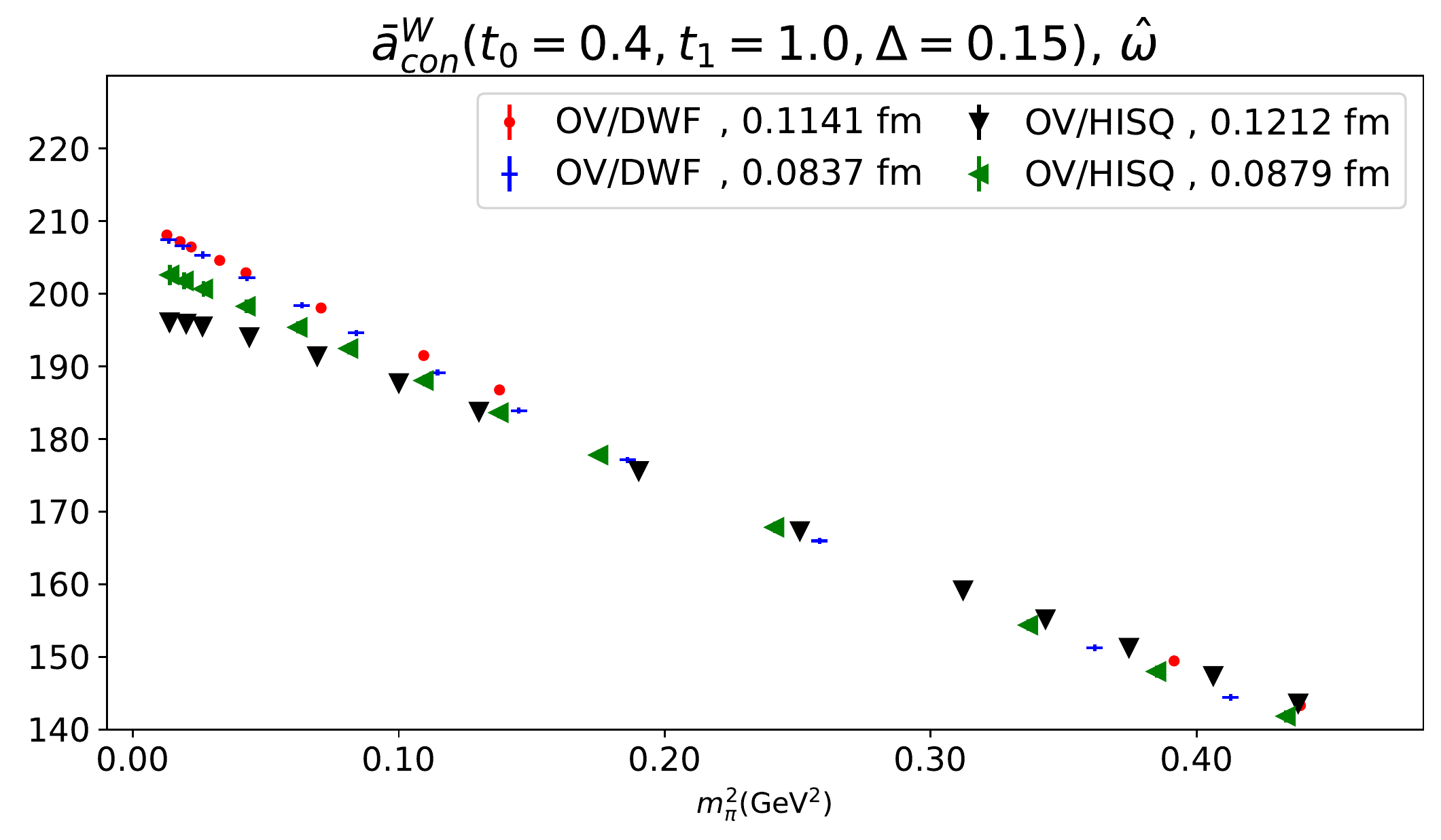}
    \caption{
    The $m_\pi^2$ dependence of the connected insertion contribution $\bar{a}^{\rm W}_{\rm con}(m_\pi^2)$ based on the original $\omega$ (defined in Eq.~\ref{eq:def_omega}, upper panel) and the modified one ($\hat{\omega}$ as defined in Eq.~\ref{eq:def_omega2}, lower panel), on four ensembles we used.}
    \label{fig:quark_mass}
\end{figure}

Since suppressing the statistical uncertainty using the bounding method for the long distance contribution can be nontrivial, we will concentrate on the medium range contribution, $\bar{a}^{\rm W}_{\rm con}$, and the short range one, $\bar{a}^{\rm S}_{\rm con}$, in this work.
In Fig.~\ref{fig:quark_mass}, we show the result of $\bar{a}^{\rm W}_{\rm con}(m_\pi^2)$ on the four physical-point ensembles as a function of the pion mass squared.
Compared to the values in the upper panel which use the original $\omega$, those in the lower panel using the discretized $\hat{\omega}$ have smaller differences between ensembles except for the light quark mass region of the OV/HISQ case. It suggests that the modified definition might suppress the discretization error.
But the $m_\pi^2$ dependence in the small $m_\pi^2$ region is nonlinear and then the difference between the OV/HISQ results from the a12m130 and a09m130 ensembles is larger with $\hat{\omega}$ compared to that using $\omega$. 

\begin{figure}[t]
    \centering
   \includegraphics[width=1\linewidth]{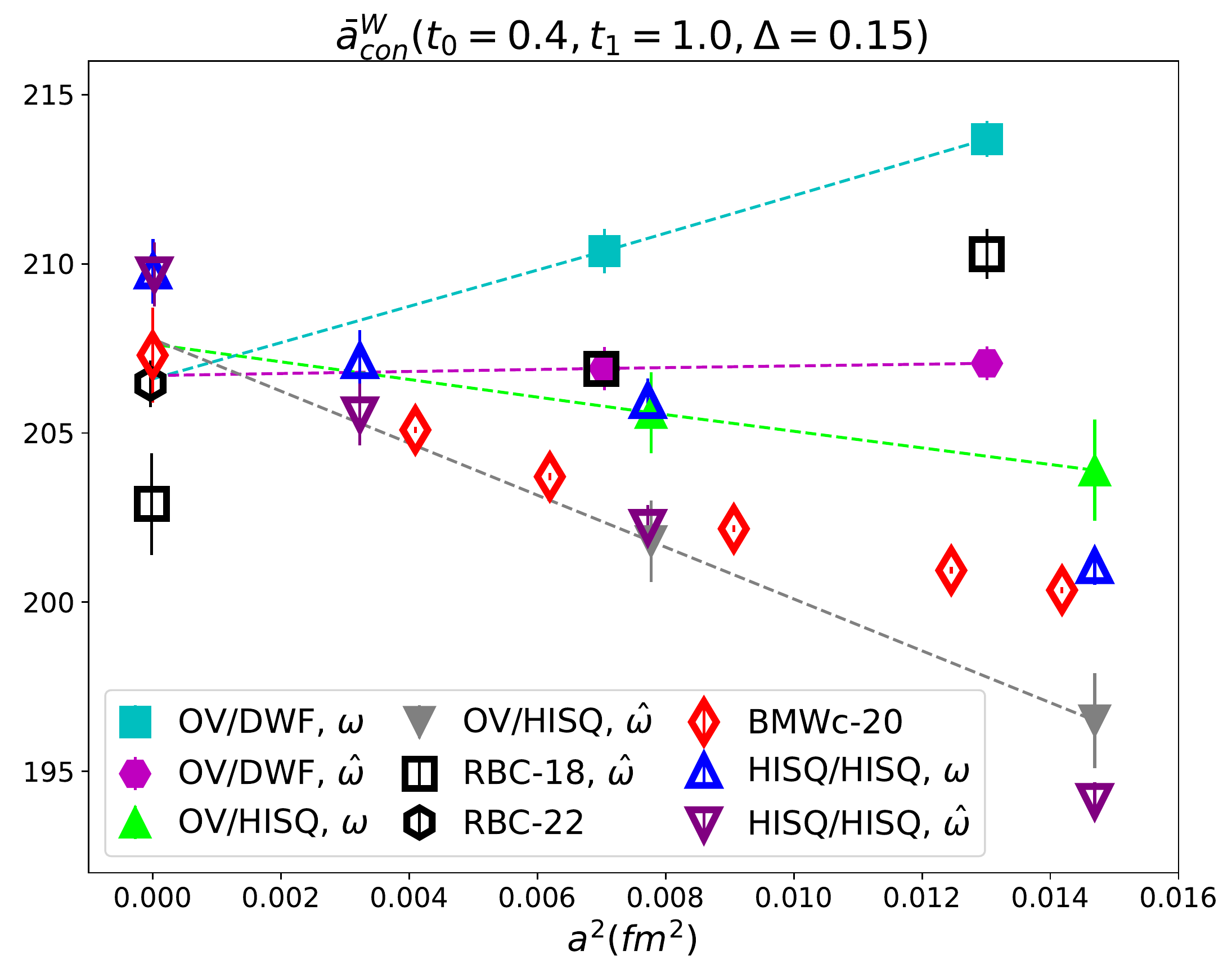}
    \caption{The lattice spacing dependence of $\bar{a}^{\rm W}_{{\rm con}, l}$ for different actions and weight functions are plotted with filled symbols. Results with unitary valence and sea actions, namely, RBC-18~\cite{RBC:2018dos} and RBC-22~\cite{RBCUKQCD:2022,Blum:2023qou} with DWF (black open boxes and black open hexagon, respectively), Aubin-19~\cite{Aubin:2019usy} with HISQ (blue and purple open triangles) and BMWc-20~\cite{Borsanyi:2020mff} with a 4stout, staggered fermion action (red open diamonds), are also shown in the figure for comparison.  The dashed lines are $a^2$ extrapolations of the mixed action results based only on two lattice spacings. }
    \label{fig:ensembles}
\end{figure}

By interpolating the partially quenched valence pion mass to the physical value 135 MeV, we obtain the light quark contribution $\bar{a}^{\rm W}_{{\rm con}, l}$ as shown in Fig.~\ref{fig:ensembles}. We note that the modification of $\omega$ to $\hat{\omega}$ suppresses the discretization error in the OV/DWF setup (cyan and purple) but has the reversed effect on the OV/HISQ setup (green and gray).
For comparison, we also show the results for unitary DWF~\cite{RBC:2018dos} (open black boxes) and unitary HISQ~\cite{Aubin:2019usy}  {(open triangles)}. 
One can see that while those at larger lattice spacings have obvious differences, our results at $a\sim$ 0.08--0.09~fm are consistent with the unitary results within uncertainties.
Note that we used the local vector current with normalization while the unitary HISQ result~\cite{Aubin:2019usy} used the conserved current, and so the agreement here could be accidental.
Such a difference would be a discretization effect since it decreases with smaller lattice spacing.
The OV/DWF and OV/HISQ results are conspicuously different at $a\sim$ 0.08--0.09~fm.
This is an indication that there is still large sea fermion action dependence at this lattice spacing.
One possible source of the OV/DWF-OV/HISQ discrepancy at $a \sim$ 0.08--0.09 fm could be related to the gauge actions used in the DWF and HISQ ensembles, as different improvements make the bare gauge coupling in the RBC ensembles (2.13--2.25) and MILC ensembles (3.60--3.78) differ by a factor of $\sim 1.7$. Such a possibility can be checked with HISQ+Iwasaki and DWF+Iwasaki simulations at $a = 0.08$ fm on a smaller lattice.

After the linear $a^2$ continuum extrapolation, the OV/DWF result using  $\hat{\omega}$, 206.7(1.5), is consistent with that using $\omega$, 206.4(1.5). Similar consistency is also found in the OV/HISQ case, with 207.7(3.1) using $\hat{\omega}$ and 207.6(3.1) using $\omega$. All the values are consistent with each other within their uncertainties.
Thus we combine these values to predict $\bar{a}^{\rm W}_{{\rm con}, l}=206.7(1.5)(1.0)$ using the OV/DWF value with $\hat{\omega}$ and smaller statistical uncertainty as the central value, and the difference of the results as a systematic uncertainty. These results are consistent with the Budapest-Marseille-Wuppertal collaboration (BMWc) value {207.3(1.4)}~\cite{Borsanyi:2020mff} and the latest RBC results~\cite{RBCUKQCD:2022,Blum:2023qou}, but are less than 2 $\sigma$ higher than the RBC-18 value 202.9(1.5)~\cite{RBC:2018dos}.

Note that ${\cal O}(a^4)$ behavior in OV/DWF and OV/HISQ has been observed in $\Delta_{mix}$, the leading order low-energy constant of the chiral perturbation theory with different valence and sea actions~\cite{Zhao:2022bke,Zhao:2021}.
Thus it's possible that the current agreement under simple linear $a^2$ continuum extrapolation may be due to some unknown cancellation of the higher-order terms under our mixed-action setups.
Various recent studies~\cite{Aubin:2022hgm,Ce:2022kxy,FermilabLattice:2022smb,Alexandrou:2022amy, RBCUKQCD:2022,Blum:2023qou} have shown that ${\cal O}(a^4)$ corrections are important, so it would be natural to further extend our studies to smaller $a$ to have better control of the continuum extrapolations.

\begin{figure}[t]
    \centering
    \includegraphics[width=1\linewidth]{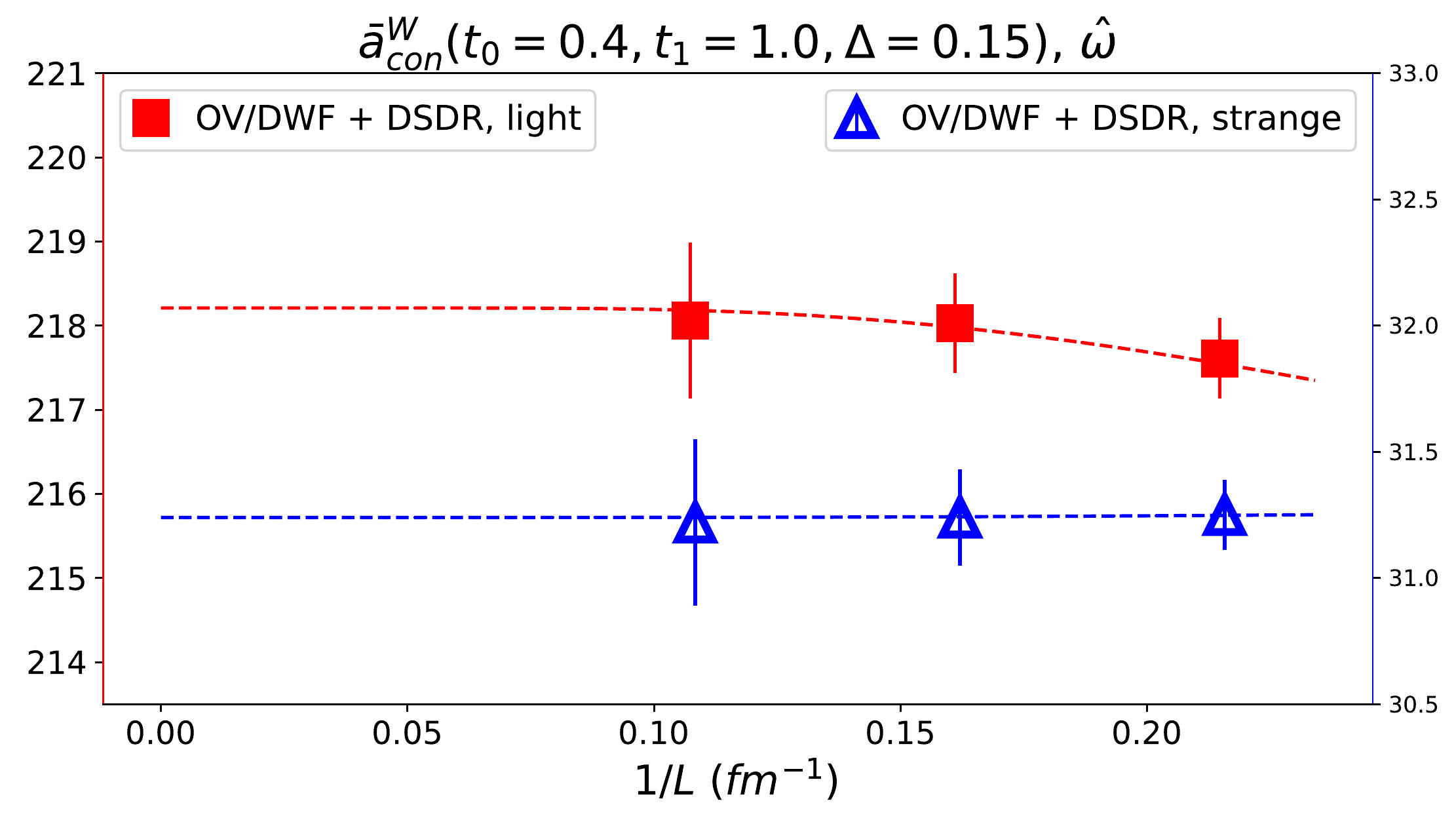}
    \caption{The volume dependence of $\bar{a}^{\rm W}_{{\rm con}, l}$ (red boxes and left $y$-axis) and $\bar{a}^{\rm W}_{{\rm con}, s}$ (blue triangles and right $y$-axis) based on the DSDR ensembles at $a=0.194~\mathrm{fm}$. The $\bar{a}^{\rm W}_{{\rm con}, l}$ values here are much larger than those in Fig.~\ref{fig:ensembles} due to the discretization errors.}
    \label{fig:volume}
\end{figure}

Since the volumes of the 4 ensembles are close to each other ($1/L\in[0.172,0.184] ~\mathrm{fm}^{-1}$), we use the Mobius+Iwasaki+DSDR ensembles~\cite{RBC:2012cbl,Boyle:2015exm} from the RBC/UKQCD collaboration to estimate the finite volume effect.
As shown in Fig.~\ref{fig:volume}, the finite volume effect with an empirical form $a+b\,\mathrm{exp}(-m_{\pi}L)$ for the case with $1/L\sim 0.18~\mathrm{fm}^{-1}$ is $-0.36(56)$ for the light quarks (red boxes, left $y$-axis) and 0.01(18) for the strange quark (blue triangles, right $y$-axis).

\begin{figure}[t]
    \centering
   \includegraphics[width=1\linewidth]{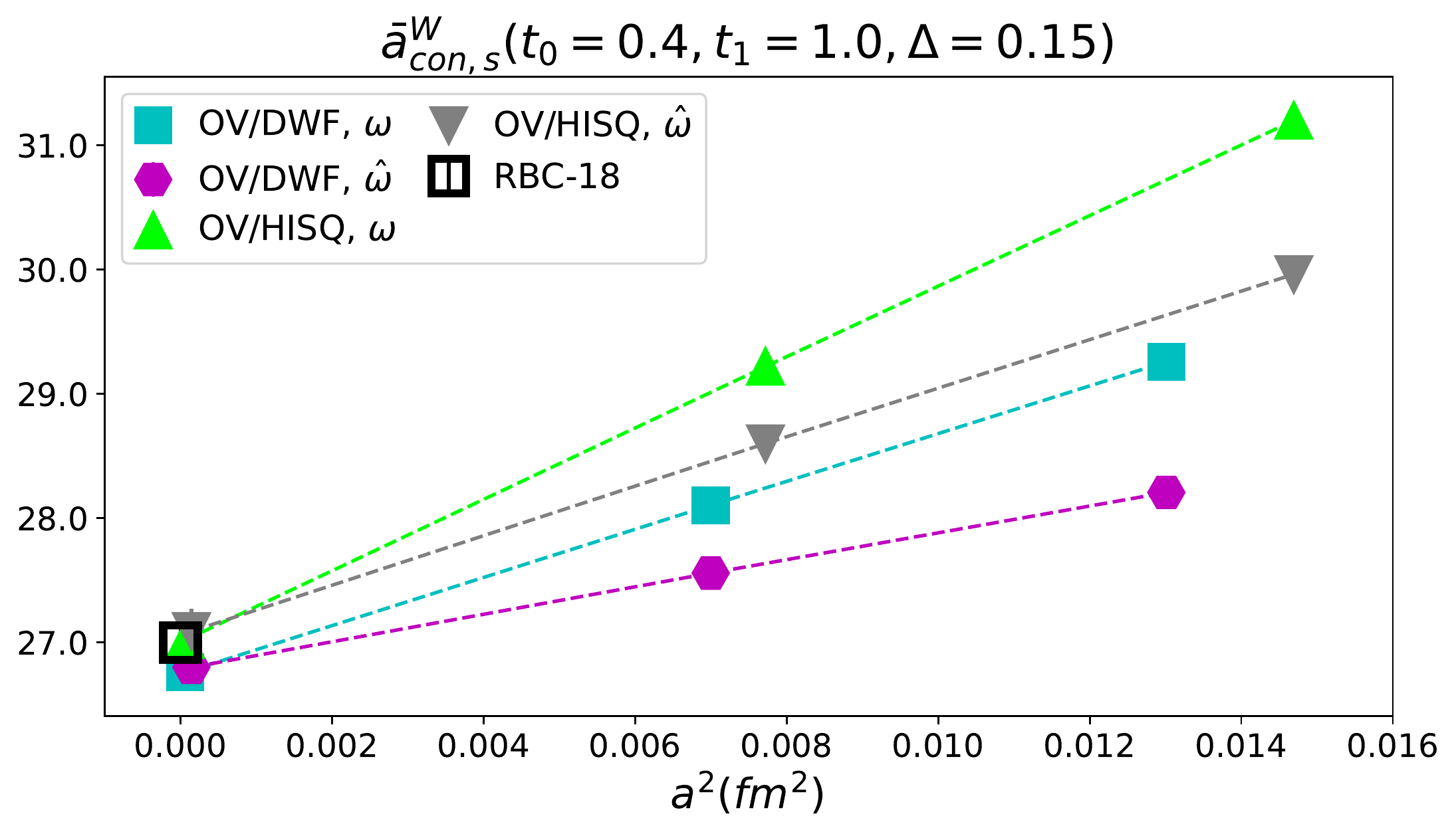}
    \caption{The lattice spacing dependence of $\bar{a}^{\rm W}_{{\rm con}, s}$. The DWF result extrapolated to the continuum limit~\cite{RBC:2018dos} is also shown in the figure for comparison.}
    \label{fig:ensembles_s}
\end{figure}

With the ratio $m_s/m_{l}$ = 27.42(12) from the FLAG review~\cite{FLAG:2019iem} and the bare valence light quark mass corresponding to the physical pion mass, we can estimate the physical bare valence strange quark mass on each ensemble. 
With this strategy, we show the strange quark contribution $\bar{a}^{\rm W}_{{\rm con}, s}$ in Fig.~\ref{fig:ensembles_s}. Similar to the light quark case, the linear $a^2$ extrapolated OV/DWF result of 26.8(1) and OV/HISQ result of 27.1(2) are consistent within the systematic uncertainty due to the strange quark mass (which is about 0.1). 
They are consistent with each other under uncertainty and also consistent with the RBC value of 27.0(2) (open black box in the figure).
At the same time, if we use the so-call $\eta_s$ ``mass" 689.89(69) MeV~\cite{Borsanyi:2020mff} to determine the physical strange quark mass, we get
OV/DWF result 26.7(3) and OV/HISQ result 26.7(6). Since the scale setting uncertainty enters the strange quark mass definitions, these results are consistent with those using the quark mass ratio $m_s/m_{l}$ but have larger uncertainties.
Thus we combine these values to predict $\bar{a}^{\rm W}_{{\rm con}, s}=26.8(1)(3)$ with the OV/DWF value using the bare quark mass ratio $m_s/m_{l}$ = 27.42(12) and smaller statistical uncertainty as the central value, and the difference of the results as a systematic uncertainty.

\begin{figure}[t]
    \centering
    \includegraphics[width=1\linewidth]{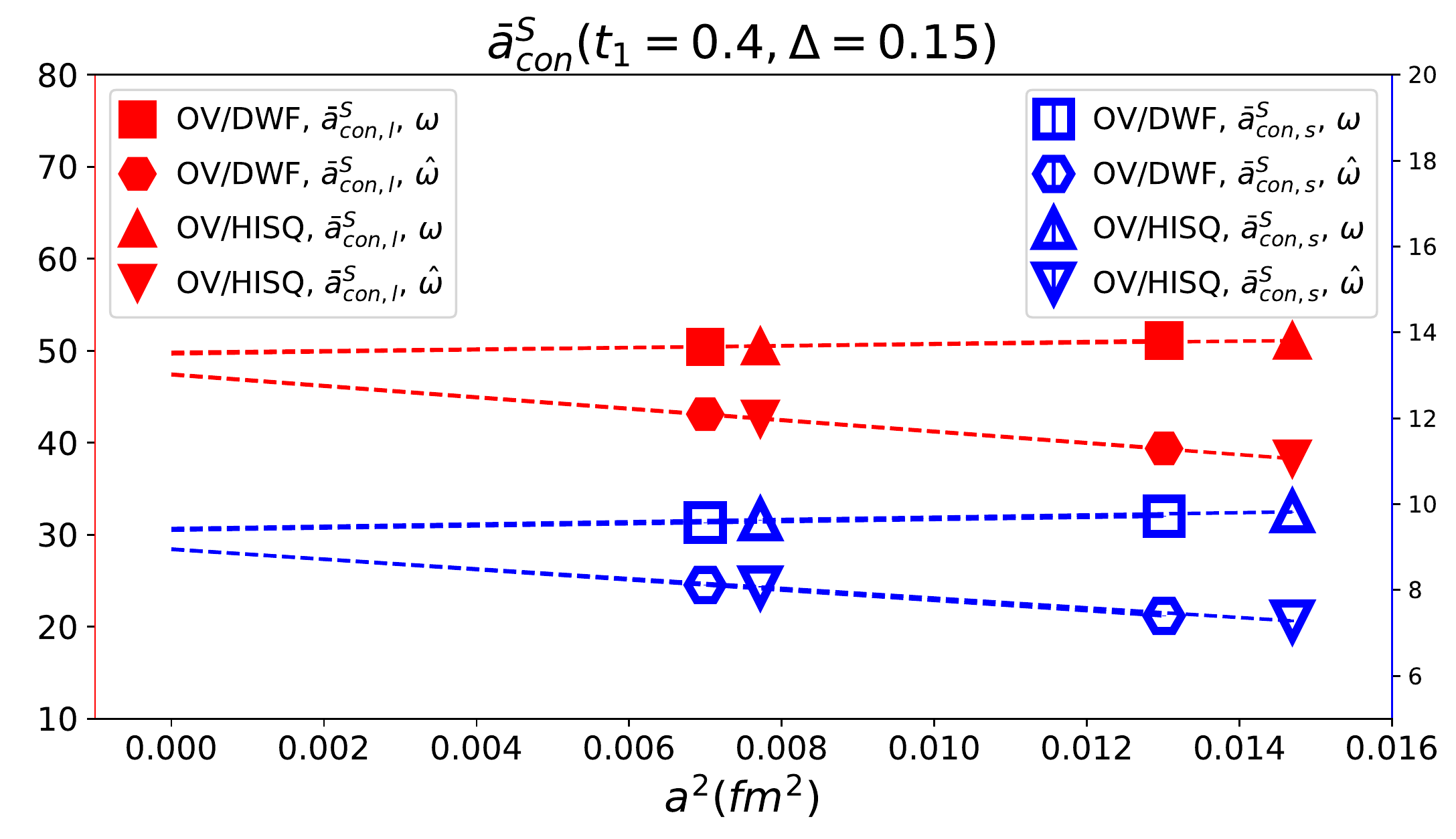}
    \caption{The lattice spacing dependence of the short distance contributions $\bar{a}^{\rm S}_{{\rm con}, l}$ (red data points with the left $y$-axis) and $\bar{a}^{\rm S}_{{\rm con}, s}$  (blue data points with the right $y$-axis), using either $\omega$ or $\hat{\omega}$. One can see that the action dependence almost vanishes and then is much weaker than that of the medium distance contribution $\bar{a}^{\rm W}_{\rm con}$.}
    \label{fig:short}
\end{figure}

Next, we turn to the short distance contribution $\bar{a}^{\rm S}_{{\rm con}, l/s}$, with
the results shown in Fig.~\ref{fig:short}. We can see that the linear ${\cal O}(a^2)$ lines are almost the same on both the RBC and MILC ensembles for both the light and strange quark mass cases, and the linear $a^2$ continuum extrapolated values of the OV/DWF and OV/HISQ setups are consistent within the uncertainty (except for the case of $\bar{a}^{\rm S}_{{\rm con}, s}$ using $\omega$ where the extrapolated values from the two setups differ by 0.04(2)).
But it is interesting that the discretization error of $\bar{a}^{\rm S}_{{\rm con}}$ using $\omega$ is much smaller compared to that using $\hat{\omega}$, and the extrapolated values using the two definitions are separated by more than $\sim$5\% difference for both the light and strange quark cases.
Thus we predict $\bar{a}^{\rm S}_{{\rm con}, l}=48.58(0.07)(1.20)$ and $\bar{a}^{\rm S}_{{\rm con}, s}=9.18(01)(25)$ with the systematic error estimated from half of the difference between the predictions of the results from $w$ and $\hat{w}$.

\begin{figure}[t]
    \centering
    \includegraphics[width=1\linewidth]{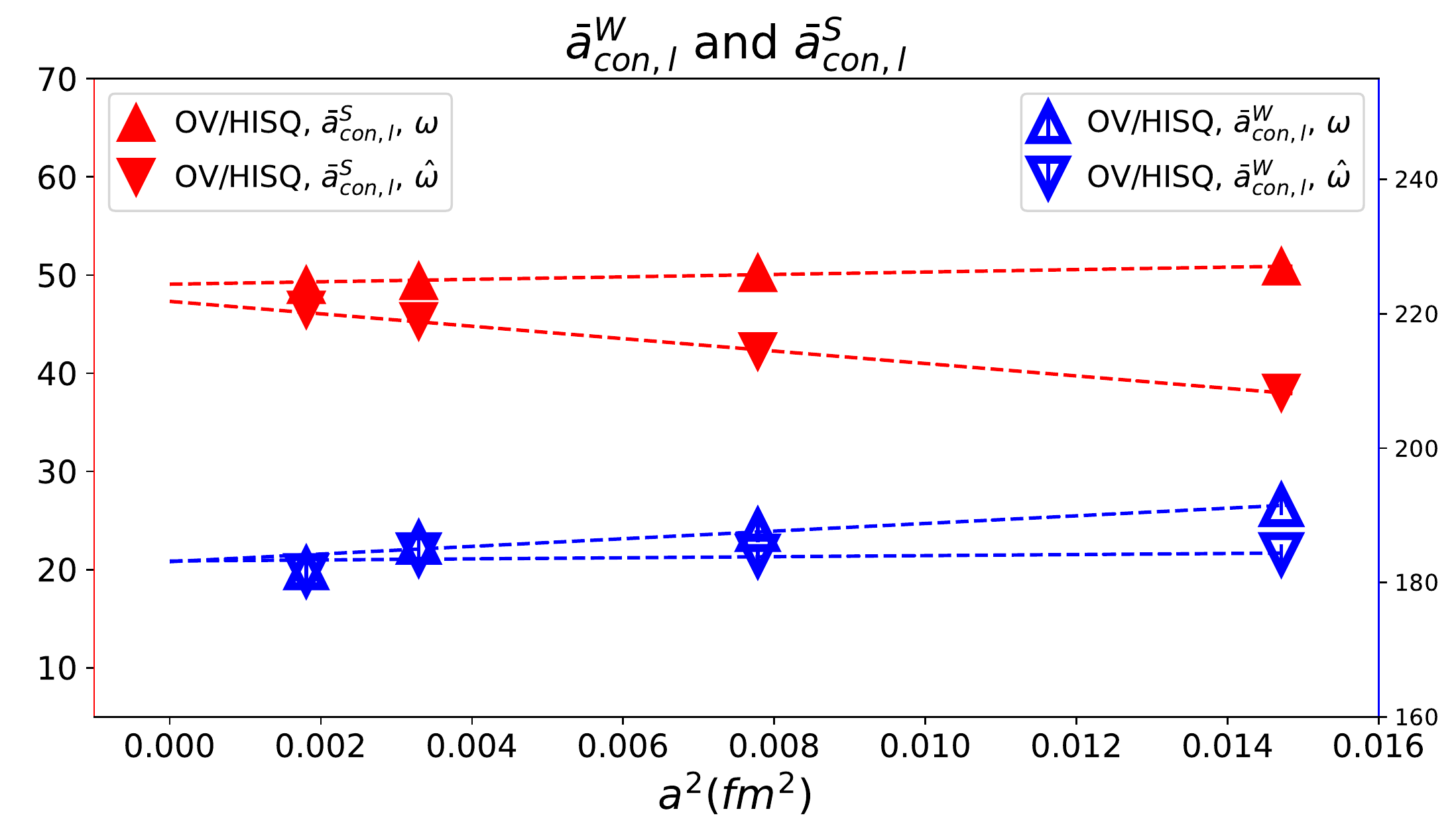}
    \caption{The lattice spacing dependence of $\bar{a}^{\rm S}_{{\rm con},l}$ (red data points with the left $y$-axis) and $\bar{a}^{\rm W}_{{\rm con},l }$  (blue data points with the right $y$-axis) on HISQ ensembles with pion mass $m_{\pi}\simeq 310$~MeV, using either $\omega$ or $\hat{\omega}$. }
    \label{fig:HISQ_heavy}
\end{figure}

This motivates us to repeat the calculation on the HISQ ensembles at a $\sim 310$~MeV pion mass with a larger lattice spacing range $a\in[0.04,0.12]$~fm to check the lattice spacing dependence. Fig.~\ref{fig:HISQ_heavy} shows that the linear $a^2$ extrapolation still works well for $\bar{a}^{\rm W}_{{\rm con},l }$ (blue data points with the right $y$-axis), and using $\hat{\omega}$ can suppress the discretization error (similar to the OV/DWF results at the physical point). On the other hand, we can also see that $\bar{a}^{\rm S}_{{\rm con},l }$ (red data points with the left $y$-axis) is less sensitive to the lattice spacing when we use the original $\omega$ instead of $\hat{\omega}$, and the tension between the linear $a^2$ extrapolated values using either $\omega$ or $\hat{\omega}$ still exists. It suggests that using $\hat{\omega}$ introduces an extra discretization error in the small $t$ region. Adding $a^4$ terms in the continuum extrapolation of $\bar{a}^{\rm S}_{{\rm con},l}$ using $\hat{\omega}$ can suppress the inconsistency.

\begin{figure}[t]
    \centering
    \includegraphics[width=1\linewidth]{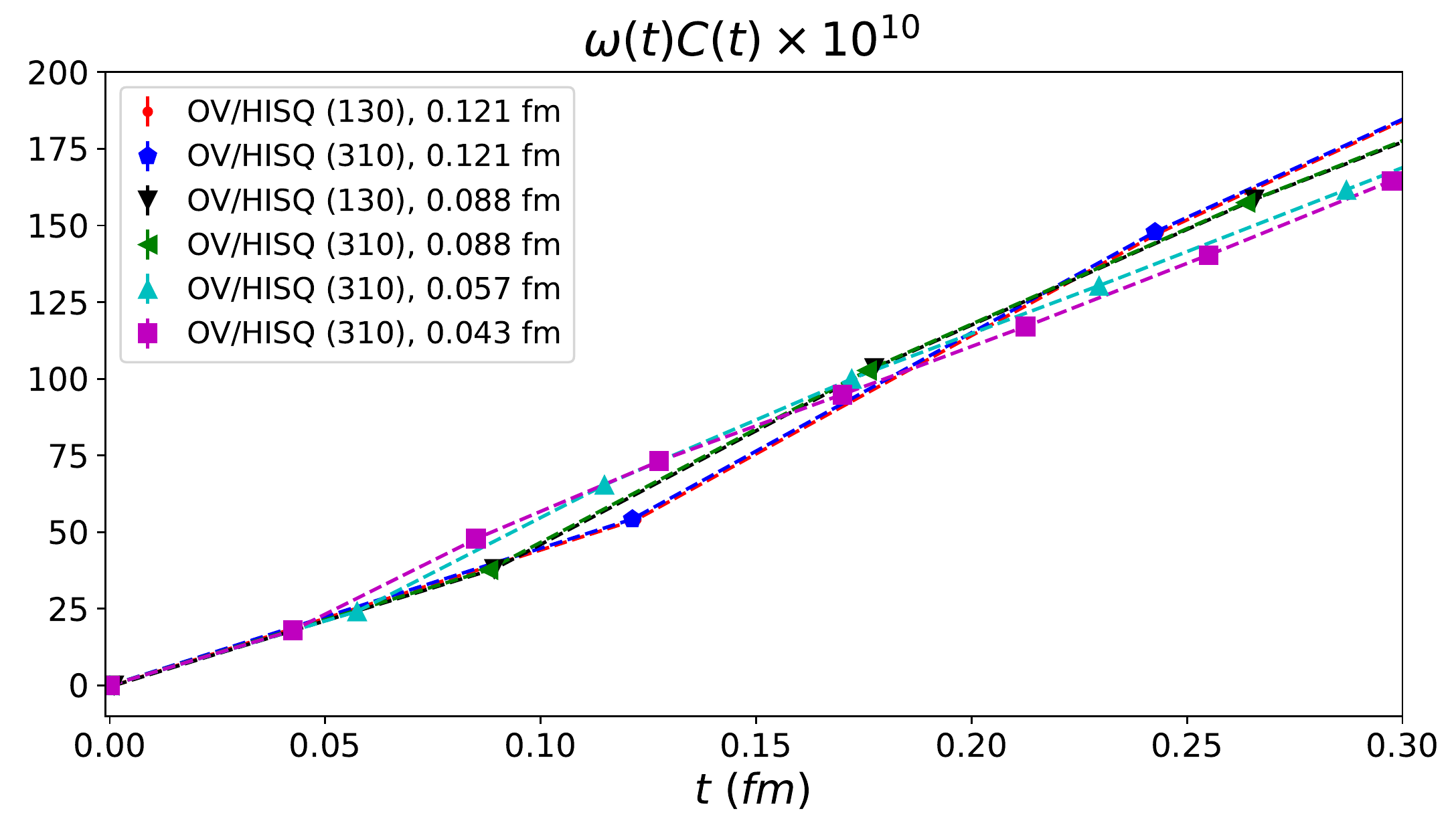}
   \includegraphics[width=1\linewidth]{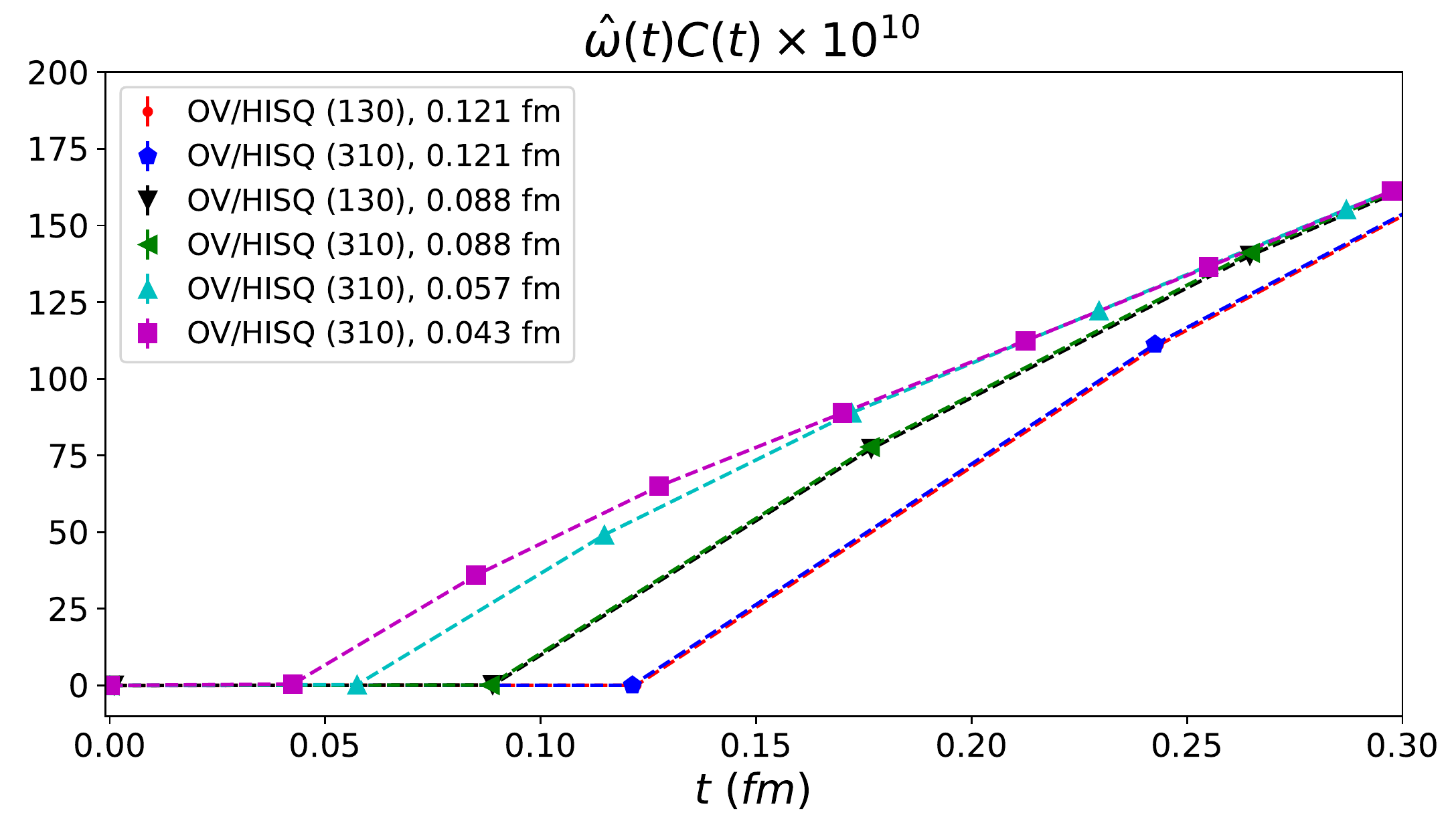}
    \caption{The values of $\omega(t)C(t)$ (upper panel) and $\hat{\omega}(t)C(t)$ (lower panel) at the small $t$ region with pion mass $m_{\pi}\simeq 310$~MeV and 135~MeV.}
    \label{fig:small_t}
\end{figure}

$\bar{a}^{\rm S}_{{\rm con},l }$ corresponds to the integral of $\omega(t)C(t)$ in the small $t$ region, and thus its discretization effect can be illustrated through the values of $\omega(t)C(t)$ (Fig.~\ref{fig:small_t}, upper panel) and those of $\hat{\omega}(t)C(t)$ (lower panel)).
The definition of $\hat{\omega}$ forces $\hat{\omega}(a)\propto a^2$, and, as a consequence, the integral of $\hat{\omega}(t)C(t)$ has a sizable discretization error around $t\sim 1a$~\cite{RBC:2018dos}.
This is illustrated in the lower panel of Fig.~\ref{fig:small_t}.
Since $\hat{\omega}(t)C(t)$ is not linear in $t$ in the range of $t\in[0.0,0.2]$~fm, the extra ${\cal O}(a^4)$ effect is not avoidable and cannot be mocked up with a linear $a^2$ continuum extrapolation.

As shown in Figs.~\ref{fig:ensembles_s} and ~\ref{fig:short}, $\bar{a}^{\rm S}_{{\rm con},l }$ is insensitive to the quark mass. It is also verified in the recent study~\cite{RBCUKQCD:2022,Blum:2023qou} and used to suppress the uncertainty of $\bar{a}^{\rm S}_{{\rm con},l }$ by combining the $\sim$ 300 MeV result at fine lattice spacing and the mass correction at larger lattice spacing. Thus we also show the values using the physical ensembles a09m130 and a12m130 in Fig.~\ref{fig:small_t}, and they are consistent with the data using heavier pion mass at short distance.

\begin{figure}[t]
    \centering
    \includegraphics[width=1\linewidth]{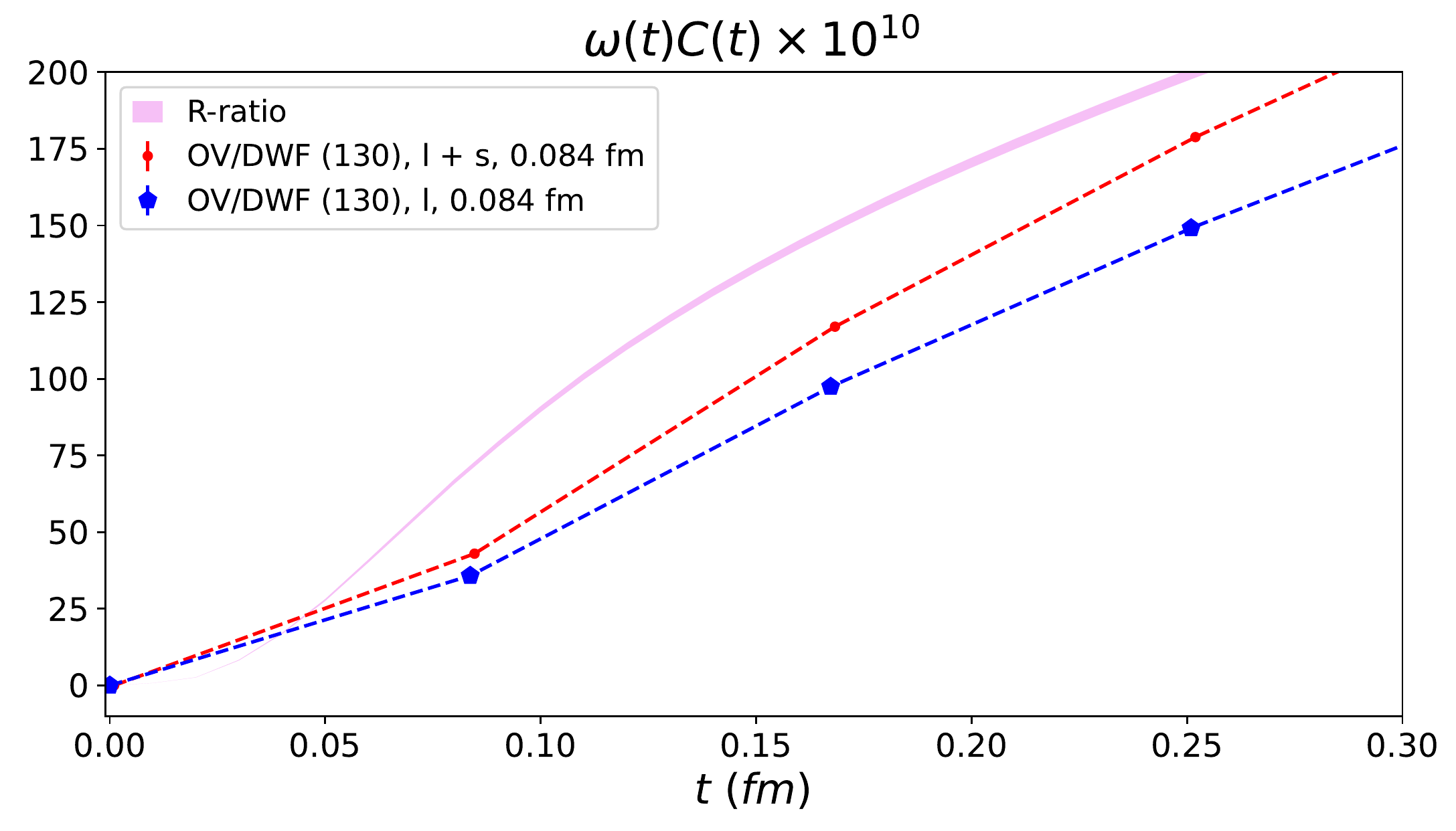}
    \caption{The values of $\omega(t)C(t)$ of the connected light and strange contributions at the small $t$ region with pion mass $m_{\pi}\simeq 130$~MeV at lattice spacing 0.084 fm.}
    \label{fig:small_t_fin}
\end{figure}

Eventually, we show the values
of $C(t)\omega(t)$ (pink band), with $C(t) = 1/(12 \pi^2) \int_0^{\infty} d(\sqrt{s}) R(s) s e^{- \sqrt{s} t}$ from ~\cite{Bernecker:2011gh} in Fig.~\ref{fig:small_t_fin}, using the most recent analysis of the $R$-ratio data~\cite{Keshavarzi:2019abf}, and compare with the connected light contribution $\frac{5}{9}C^{\rm con}(t,m_l)\omega(t)$ (blue dots) and the connected light+strange contribution $(\frac{5}{9}C^{\rm con}(t,m_l)+\frac{1}{9}C^{\rm con}(t,m_s))\omega(t)$ (red dots) based on the OV/DWF result at 0.084 fm. The pink band is about 40\% higher than the blue dots at $t\sim 0.1$~fm while the relative difference becomes smaller at $t\sim 0.2$~fm. The difference should be primarily due to the connected charm contribution and is worth further investigation in the future.

\section{Summary and Discussion}\label{sec:summary}

In this work, we calculated the light and strange contributions of $a_{\mu}$ from the connected vector correlators in the medium window ($t_0=0.4~\mathrm{fm}$, $t_1=1.0~\mathrm{fm}$, $\Delta=0.15~\mathrm{fm}$) using the overlap fermions, on the physical point ensembles using either DWF+Iwasaki (at $a=0.084/0.114$~fm) or HISQ+Symanzik (at $a=0.088/0.121$~fm) configurations. Our linear $a^2$ extrapolated $\bar{a}_{{\rm con}, s}^{\rm W}$ result is 26.7(3) using the OV/DWF setup; it is consistent with the value 27.5(6) using the OV/HISQ setup and also with the unitary DWF value from RBC~\cite{RBC:2018dos}.

For $\bar{a}_{{\rm con}, l}^{\rm W}$, the mixed action results on the ensembles with $a<0.1$ fm are consistent with the unitary DWF or HISQ calculations, but those at $a>0.1$ fm are different from their respective unitary results by many sigmas.
After linear $a^2$ continuum extrapolations, the OV/DWF and OV/HISQ results are consistent with each other and are combined to give 206.7(1.5)(1.0) which is consistent with the BMWc value 207.3(1.4)~\cite{Borsanyi:2020mff} and with the recent RBC update~\cite{RBCUKQCD:2022}. Furthermore, we note that using $\hat{\omega}$ can suppress the discretization error of $\bar{a}_{{\rm con}, l}^{\rm W}$ with the OV/DWF setup, but this is not the case with the OV/HISQ setup. Such an observation is similar to that using the unitary DWF or HISQ setups~\cite{RBC:2012cbl,Aubin:2019usy}.

We also calculated the short range contribution and predict $\bar{a}^{\rm S}_{{\rm con},l+s}=57.8(0.1)(1.5)$ with the systematic error estimated from half of the difference between the predictions of the results from $\omega$ and $\hat{\omega}$. Such a systematic uncertainty is much larger than the statistical uncertainty, as the result using $\hat{\omega}$ has a much stronger discretization error than that using $\omega$. Our result also shows that sensitivity to the sea fermion action is much weaker than that in the window range. It would be valuable to verify our observations on other lattice setups.

Based on our calculation in both the short distance and window ranges we find that the sensitivity to the sea fermion (or gauge) action and to using either $\hat{\omega}$ or $\omega$ is range dependent. Thus a similar study of the long distance range contribution should be important to improve our understanding of the discretization error there. We also suggest a unitary HISQ+Iwasaki simulation at around $a=0.09$ fm to test the gauge action dependence.

\section*{Acknowledgments}
We thank the RBC/UKQCD, and MILC collaborations for providing us their gauge configurations, A. Keshavarzi for sharing their $R$-ratio data, and F. He, C. Lehner and L. Lellouch for valuable inputs and discussion. K.L.\ thanks T. Blum and A. El-Khadra for the informative discussions. Most of the production was performed using the the GWU code~\cite{Alexandru:2011ee,Alexandru:2011sc} through the HIP programming model~\cite{Bi:2020wpt}, and the data analysis was based on the Qlattice package. T.D. and K.L. are supported in part by the U.S. DOE Grant No.\ DE-SC0013065 and K.L. by DOE Grant No.\ DE-AC05-06OR23177, which is within the framework of the TMD Topical Collaboration. G.W. is supported by the French National Research Agency under the contract ANR-20-CE31-0016. Y.Y.\ is supported in part by a NSFC-DFG joint grant under Grants No.\ 12061131006 and SCHA 458/22 and also the Strategic Priority Research Program of Chinese Academy of Sciences, Grant No.\ XDC01040100, XDB34030303 and XDPB15. 
The numerical calculations were carried out on the ORISE Supercomputer, and HPC Cluster of ITP-CAS. This research used resources of the Oak Ridge Leadership Computing Facility at the Oak Ridge National Laboratory, which is supported by the Office of Science of the U.S. Department of Energy under Contract No.\ DE-AC05-00OR22725. This work used Stampede time under the Extreme Science and Engineering Discovery Environment (XSEDE), which is supported by National Science Foundation Grant No.\ ACI-1053575.
We also thank the National Energy Research Scientific Computing Center (NERSC) for providing HPC resources that have contributed to the research results reported in this paper.
We acknowledge the facilities of the USQCD Collaboration used for this research in part, which are funded by the Office of Science of the U.S. Department of Energy. 

\bibliographystyle{apsrev4-1}
\bibliography{reference.bib} 

\end{document}